\documentclass[namedreferences]{solarphysics}

\usepackage[applemac]{inputenc} 
\usepackage[hyperref,optionalrh,solaromanenum]{spr-sola-addons} % For Solar Physics 
\usepackage{graphicx}                    % For eps figures, newer & more powerfull
\usepackage{color}                       % For color text: \color command
\usepackage{breakurl}                         % For breaking URLs easily trough lines
\begin{document}

\begin{article}

\begin{opening}

\title{Christian Horrebow's Sunspot Observations -- II. Construction of a Record of Sunspot Positions}

%%%%%%%%%%%%%%%%%%%%%%%%%%%%%%%%%%%%%%%%%%%%%%%%%%%
%% Authors Names
%
% \author[addressref={},corref,email={}]{\inits{}\fnm{}\lnm{}\orcid{}}
 \author[addressref={aff2,aff1},corref,email={karoff@phys.au.dk}]{\inits{C.}\fnm{Christoffer } \lnm{Karoff $^{*,}$}}\orcid{https://orcid.org/0000-0003-2009-7965}
 \author[addressref={aff1}]{\inits{C.S.}\fnm{Carsten Sønderskov } \lnm{Jørgensen}}\orcid{https://orcid.org/0000-0002-0319-1207}
 \author[addressref={aff3}]{\inits{V.S.P.}\fnm{Valliappan Senthamizh } \lnm{Pavai}}\orcid{https://orcid.org/0000-0003-2413-7901}
 \author[addressref={aff3},email={rarlt@aip.de}]{\inits{R.}\fnm{Rainer }\lnm{Arlt}}

%%%%%%%%%%%%%%%%%%%%%%%%%%%%%%%%%%%%%%%%%%%%%%%%%%%
%% Runningheads
%
%\runningauthor{}
%\runningtitle{}

%%%%%%%%%%%%%%%%%%%%%%%%%%%%%%%%%%%%%%%%%%%%%%%%%%%
%% Affilations 
%% id shold be the same with \author addressref value.
%\address[id={}]{}
\address[id=aff1]{Stellar Astrophysics Centre, Department of Physics and Astronomy, Aarhus University, Ny Munkegade 120, 8000, Aarhus C, Denmark}
\address[id=aff2]{Department of Geoscience, Aarhus University, H{\o}egh-Guldbergs Gade 2, 8000, Aarhus C, Denmark}
\address[id=aff3]{Leibniz-Institut für Astrophysik Potsdam, An der Sternwarte 16, 14482 Potsdam, Germany}
%\address[id={}]{}
%%%%%%%%%%%%%%%%%%%%%%%%%%%%%%%%%%%%%%%%%%%%%%%%%%%
%%% Abstract 
\begin{abstract}
The number of spots on the surface of the Sun is one of the best tracers of solar variability we have. The sunspot number is not only known to change in phase with the 11-year solar cycles, but also to show variability on longer time scales. It is however, not only the sunspot number that changes in connection with solar variability. The location of the spots on the solar surface is also known to change in phase with the 11-year solar cycle. This has traditionally been visualised in the so-called butterfly diagram, but this is only well constrained form the beginning of the 19th century. This is unfortunately, as knowledge about the butterfly diagram could aid our understanding of the variability and the Sun-Earth connection.

As part of a larger review of the work done on sunspots by the Danish astronomer Christian Horrebow, we here present a reanalysis of Christian Horrebow's notebooks covering the years 1761 and 1764--1777. These notebooks have been analysed in at least three earlier studies by Thiele (1859, Astronomische Nachrichten 50, 257), d'Arrest (published in Wolf, 1873, Astronomische Mitteilungen der Eidgen{\"o}ssischen Sternwarte Zurich 4, 77) and Hoyt and Schatten (1995, Solar Phys. 160, 387). In this article, we construct a complete record of sunspot positions covering the years 1761 and 1764--1777. The resulting butterfly diagram shows the characteristic structure known from observations in the 19th and 20th century. We do see some indications of equatorial sunspots in the observations we have from Cycle 1. However, in Cycle 2, which has much better coverage, we do not see such indications.
\end{abstract}

%%%%%%%%%%%%%%%%%%%%%%%%%%%%%%%%%%%%%%%%%%%%%%%%%%%
%% Keywords
%
%\keywords{}

\end{opening}

\section{Introduction}
Christian Pedersen Horrebow was director of the Royal Danish Observatory at Rundetårn in Copenhagen (55$^{\circ}$40$'$53.00$''$N, 12$^{\circ}$34$'$32.99$''$E) from 1753 until his death in 1776. In \citet{Jorgensen18} (Paper I) we present a biography and the complete bibliography of Christian Horrebow. We also present English translations of 11 of Christian Horrebow's publications in {\it Dansk Historisk Almanak} and one in {\it Skrifter som udi Kiøbenhavnske Selvskab af Lærdoms og Videnskabers Elsker}. 

Based on these publications we argue that Christian Horrebow was more productive than previously given credit for with detailed observations being performed over more than a decade under his leadership. Already in 1775 Christian Horrebow thus suggested that the Sun repeats itself with respect to the number and size of spots after a certain number of years, i.e. that the Sun is cyclic. We suggest that the reason why Christian Horrebow's finding has not attracted more attention is due to the tarnished reputation he earned due to the unfortunate affair with the 1761 Venus transit and because Christian Horrebow published his findings in Danish.

A similar phenomenon is seen with the discovery of the butterfly diagram or Spoerer's law, which is generally credited to Richard Carrington. Malapert however, noted the jump from near-equator to high-latitude sports at the cycle transition already in 1620 \citep{2016AN....337..581N}.

Here, we continue and analyse the observing notebooks made under Christian Horrebow. These notebooks have previously been analysed by \citet{1859AN.....50..257T},  d'Arrest \citep[published in][]{1873MiZur...4...77W} and  \citet{1995SoPh..160..387H}, who also constructed their individual sunspot record based on them. 

Christian Horrebow made his sunspot observations at a time og which we have very few other records of sunspot observations \citep{1998SoPh..179..189H, 1998SoPh..181..491H, 2016SoPh..291.2629C, 2016SoPh..291.2653S}. In fact the solar disc drawings by German astronomer Johann Caspar Staudacher, which covers the period from 1749 to 1799, appear to be the only observations covering this period. Additionally, the observations are from a period where the different reconstructions of solar activity disagree significantly \citep{ 2016SoPh..291.2629C}.

Another particular thing about the solar cycles in the last part of the 18th century is that they deviated significantly from the mean period length of $11.0\pm1.2$ year \citep{2015LRSP...12....4H} as Cycles 3 and 4 were only 9.0 and 9.3 years, respectively, while Cycle 4, with 13.6 years, is the longest cycle on record. There is however, evidence that suggests that Cycle 4 was indeed two cycles \citep{2009ApJ...700L.154U, 2015A&A...575A..77K}. This would bring the mean cycle length closer to 10 years.

Though it could be argued that it would be appropriate to make yet another sunspot reconstruction based on Christian Horrebow's notebooks, this is not the aim of this study. Instead we want to construct a record of the sunspot positions based on Christian Horrebow's notebooks. From our analysis in Paper I and our analysis of the notebooks, it is clear that such a record can be constructed based on them. Such a record would clearly aid our understanding of the evolution of the solar cycle and the solar dynamo as it would allow us to make a butterfly diagram, measure the strength of the surface differential rotation and measure the sunspot group tilt angles. All, important ingredients in understanding the evolution of the solar cycle and in making dynamo models. 

A reconstruction of the butterfly diagram in the 18th century has been produced by \citet{2009SoPh..255..143A} \citep[see also][]{2008SoPh..247..399A} based on the drawings by Staudacher. This reconstruction suggests that equatorial sunspots were common in the 18th century, in between the Maunder and Dalton minima, especially in Cycles 0 and 1. This phenomenon has however, rarely been observed since the Dalton minimum. Equatorial sunspots suggest the presence of a dynamo mode that is symmetric around the equator \citep{2009SoPh..255..143A}. Though \citet{2009SoPh..255..143A} argued that the equatorial sunspots are not related to the quality of the observations, the finding is of such an importance that it called for an independent test. In other words, do Christian Horrebow's observations confirm the presence of equatorial sunspots in Cycle 1 and if so, have they disappeared in Cycle 2?

This article is arranged as follows. In Section 1 we provide a general introduction to Christian Horrebow and the important questions related to his work. In Section 2 we provide a more specific introduction to the notebooks. An analysis of the notebooks is presented in Section 3. The reconstruction of sunspot positions is given in Section 4 and a butterfly diagram is presented in Section 5. A discussion of our findings can be found in Section 6 and concluding remarks in Section 7.

\section{The Notebooks}
The sunspot observations from Rundetårn were documented in 36 handwritten notebooks. The observations by Christian Horrebow and his assistants for the year 1761 and 1764--1777 can be found in books 1--21. It is likely that the notebooks containing observations from before 1761 and observations for the years 1762 and 1763 were lost in the bombardment of Copenhagen in 1807. Observations from the solar eclipse in 1787 can be found in book 24 and observations from 1806 can be found in book 30. Books 22-23 and 25-29 do not contain any sunspot observations. The notebooks can be separated into two different categories: observations protocols and fair-copies. Here books, 8, 10, 12, 14, 16, and 19 fall in the category of fair-copies and there rest are protocols. Nearly all the text is in Latin, except for some technical comments in book 4 and 30, which are in Danish.
%It is however, clear that the quality of books, 1-7, 9, 11, 13, 15, 17 and 18 is generally higher than the quality of the others in the sense that the writing is easier to read, the observations are made in chronological order and that they often contain drawings. 

The quality of the drawings varies a lot, sometimes they appears as real drawings of high quality with detailed distinction between umbrae and penumbrae and sometimes they appear more as sketches supporting the tables. In general the drawings do appear as supporting material for the tables and the positions provided in the tables seem much more reliable than the positions provided in the drawings.

From time to time the notebooks contain detailed drawings of houses, woods and fields often made with the instrument {\it Rota Meridiana}. These are likely made in order to test if the instrument has changes since the previous year.  

In general, the fair-copies are systematised in a way, which suggests that these books have another purpose than the protocols. The fair-copies do also not contain a single drawing. All, expect one of the fair-copies contain the same tables as in the protocols. Book 18, which is a protocol, ends however, very strangely on 30 June 1776, even though the notebook contains a lot of blank pages hereafter. Here book 19, which is a fair-copy can be used, as it contains the entire year 1776. The sudden end of book 18 could indicate that the notebook was in fact mainly written by Christian Horrebow, as he died on 15 September 1776 though Leivog was still the most common observer after July 1775 (see Paper I).

Apart from the sunspot observations the notebooks contain stellar observations and solar observations for correcting the time. The purpose of the stellar observations was mainly to obtain a list of fixed stars that could be used for determining parallaxes. The solar observations for correcting the time, were conducted with {\it Machina Æquatorea} symmetrically before and after mid-day.

\begin{table}
\caption{Content of the different notebooks}\label{nootbooks}
\begin{tabular}{lccc}     
\hline
 & Kind & Period & Number of observations\\
 \hline
 Book 1 & Protocol & 5 May 1761 -- 28 November 1761 & 131\\
 Book 2 & Protocol & 15 January 1764 -- 26 November 1765 & 57\\
 Book 3 & Protocol & 14 February 1766 -- 31 December 1766  & 59\\
 Book 4 & Protocol & 1 January 1767 -- 31 December 1767 & 182\\
 Book 5 & Protocol & 1 January 1768 -- 24 December 1768 & 286\\
 Book 6 & Protocol & 3 January 1769 -- 18 December 1769 & 179\\
 Book 7 & Protocol & 1 January 1770 -- 18 December 1770 & 180\\
 Book 8 & Fair-copy & - & -\\
 Book 9 & Protocol & 6 January 1771 -- 19 December 1771 & 141\\
 Book 10 & Fair-copy & - & -\\
 Book 11 & Protocol & 1 January 1772 -- 23 December 1772 & 81\\
 Book 12 & Fair-copy & - & -\\
 Book 13 & Protocol & 4 January 1773 -- 31 December 1773 &115\\
 Book 14 & Fair-copy & - & -\\
 Book 15 & Protocol & 5 January 1774 -- 13 November 1774 & 78\\
 Book 16 & Fair-copy & - & -\\
 Book 17 & Protocol & 1 March 1775 -- 28 December 1775 & 108\\
 Book 18 & Protocol & 21 January 1776 -- 29 July 1776 & 113\\
 Book 19 & Fair-copy & 31 July 1776 -- 28 December 1776 & 56\\
 Book 20 & Protocol & 2 January 1777 -- 4 October 1777 & 23\\
 Book 21 & Protocol & 1 August 1777 -- 9 December 1777 & 36\\
 Book 22 & Protocol & - & 0\\
 Book 23 & Protocol & - & 0\\
 Book 24 & Protocol & 15 June 1787 & 2\\
 Book 25 & Protocol & - & 0\\
 Book 26 & Protocol & - & 0\\ 
 Book 27 & Protocol & - & 0\\
 Book 28 & Protocol & - & 0\\
 Book 29 & Protocol & - & 0\\
 Book 30 & Protocol & 21 December 1806 -- 13 October 1807 & 9\\
 \hline
\end{tabular}
\end{table}

The main instrument for sunspot observations was {\it Machina Æquatorea} (see Paper I for a detailed description of this instrument). Observations with this instrument however, first started in 1767 (book 4). In the earlier years {\it Rota Meridiana} and {\it Qvadrans} were the most used instruments and the one called {\it Machina Parallactica} was used a few times. The first part of book 4 contains a number of descriptions, often in Danish, of adjustment to {\it Machina Æquatorea} and it is clear that a lot of effort was put into improving the observations. None of the later books contain a similar description of adjustments to the instruments and so it appears that the adjustments were successful and the obtained precision satisfactory. The last description of such an adjustment is from 25 September 1767. The observations with {\it Machina Æquatorea} continued until 1777.

The observations conducted from 1767 to 1777 with {\it Machina Æquatorea} all contain a table with two columns and a drawing. As in the articles in the {\it Dansk Historisk Almanak} the tables describing the observations performed with {\it Machina Æquatorea} contain a horizontal axis with positions in hours, minutes, and seconds and a vertical axis with positions in screw turns. The positions were recorded as described in Paper 1. All observations contain measurements of the diameter of the Sun in both directions. In the vertical direction this can be used to transform the screw turns into hour angles.

The first notebooks only contain simple drawings, which seem more to play the role of illustrating the content in the tables than actually illustrating what goes on the Sun. This changes from 1770 (book 7) with the inauguration of the {\it Telescopium Gregorianum 2'}. It appears that this telescope was only used for making drawings and not for producing the content in the tables. The drawings produced with {\it Telescopium Gregorianum 2'} are however, of a much higher quality than the drawings made with {\it Machina Æquatorea} and they often contain clear distinctions between umbra and penumbra. 

It is clear that the observations were not only performed by Christian Horrebow, but also, or mainly, by a number of assistants. It is likely that Christian Horrebow wrote the articles in {\it Dansk Historisk Almanak} (see Paper I for this discussion).

\section{Analysis of the Notebooks}
The aim of the analysis of the notebooks is to produce a sunspot record containing the heliographic coordinates of the spots. From these records we will construct a butterfly diagram and in future publications we plan to analyse differential rotation and sunspot tilt angles. The aim of the analysis is not to construct the most reliable record of the sunspot number as this would demand going carefully through all the handwritten Latin comments in the notebooks.

We have not provided a record of sunspot groups in the notebooks, only of the individual sunspots. While reading the notebooks, it was not clear to us, how the sunspot groups could be counted in an objective way. We do however, speculate that the heliographic coordinates can be used, in an objective and transparent way, to calculate the number of sunspot groups. If fact, such a method could be a side product of measuring the sunspot tilt angles. We have however, not tested if this is possible. 

We have also not been able to estimate the sizes of sunspots. Most of the spots in the drawings are just point sources that do not allow the area to be measured. Though some drawings do, as these are more realistic, using them for measuring sizes of sunspots would likely be highly biased to larger spots, as they would attract more attention and therefore the observers would been much more inclined to make a realistic drawing of a larger spot, than of a small one. 

In order to construct a sunspot record containing heliographic coordinates, we reanalysed Book 1--21. 

Book 24 and 30 were omitted as observations were performed under the supervision of Thomas Bugge and not Christian Horrebow and as these observations are rather detached from the other ones and of a lower quality.

Some of Book 1, 2 and 3 only include one coordinate in the tables and no drawings. As we cannot directly calculate heliographic coordinates based on this, the observations have also been omitted (an inference of the positions similar to \citet{2018AN....339..219N} goes beyond the scope of this study). We do, however, note that the information in notebook 1, 2, and 3 is clearly good enough for calculation of the sunspot number, only the heliographic coordinates are a problem.

A number of tests with following spots on different days have shown us that the information in the tables is more reliable than the information in the drawings. Therefore, whenever informations are available in the tables we use them. It does however, happens that more spots are provided in the drawings than in the tables. We therefore want to include the information in the drawings whenever it is not available in the tables. For this we made a small program which would overlay the spots in the tables on the images of the drawings. The image size and rotation could be adjusted to align the spots from the tables with the spots in the drawings. When this was done, it was possible to mark, in the images of the drawings, the positions of spots, which are not present in the tables; those positions would then be provided in the same coordinate system as in the tables -- except for a few cases where the tables were made from the top and not the bottom limb.

The task of aligning the tables and the drawings was not trivial as the match was not perfect. We expect that this is due to the drawings being illustrations of the tables rather than realistic drawings of the Sun.

A complete table with all the processed results for all sunspots, including heliographic coordinates, can be found in the electronic supplementary material. An example of some of the information in the table can be found in Table~2.
\begin{table}
\caption{Example of sunspot position reconstruction. The complete table with additional information is available as electronic supplementary material. RM stands for {\it Rota Meridiana}.}
\begin{tabular}{lcccc}
\hline \hline
Date & Image & Longitude & Latitude & Instrument \\
\hline
1761 05 05 10 57 & 6980b & 8.8  & 7.2 & RM \\
1761 05 05 10 57 & 6980b & 3.3  & 1.1 & RM \\
1761 05 05 10 57 & 6980b & 352.6 & 19.8 & RM \\
1761 05 05 10 57 & 6980b & 283.0 & -9.6 & RM  \\
1761 05 05 10 57 & 6980b & 255.5  & 0.4 & RM  \\
1761 05 05 10 57 & 6980b & 251.7 & -2.1 & RM  \\
1761 05 05 10 57 & 6980b & 237.1 & -4.3 & RM  \\
1761 05 08 10 56 & 6984 & 322.3 & 42.7  & RM  \\
1761 05 08 10 56 & 6984 & 311.9 & 42.4  & RM  \\
1761 05 08 10 56 & 6984 & 308.6 & 39.1  & RM  \\
1761 05 08 10 56 & 6984 & 307.4 & 43.8  & RM  \\
1761 05 08 10 56 & 6984 & 300.3 & 24.6  & RM  \\
1761 05 08 10 56 & 6984 & 287.2 & -12.0  & RM \\
1761 05 08 10 56 & 6984 & 261.2 & -7.5   & RM  \\
1761 05 08 10 56 & 6984 & 257.2 & -7.2   & RM  \\
1761 05 08 10 56 & 6984 & 222.8 & 25.0  & RM  \\
1761 05 08 10 56 & 6984 & 221.5 & 24.6   & RM  \\
1761 05 08 10 56 & 6984 &  220.9 & 22.5  & RM  \\
1761 05 08 10 56 & 6984 & 218.2 & 23.1   & RM  \\
1761 05 08 10 56 & 6984 & 240.6 & -8.5    & RM  \\
1761 05 08 10 56 & 6984 & 215.8 & 20.7   & RM  \\
1761 05 08 10 56 & 6984 & 205.2 & 25.1   & RM  \\
1761 05 08 10 56 & 6984 & 231.1 & -3.7    & RM  \\
1761 05 08 10 56 & 6984 & 220.8 & -12.9   & RM \\
1761 05 08 10 56 & 6984 & 195.2 & -12.2   & RM  \\
... & ... & ... & ... & ... \\
... & ... & ... & ... & ... \\
... & ... & ... & ... & ... \\
\hline
\end{tabular}
\end{table}

\section{The Sunspot Number Series}
We have used the information in Table 2 to calculate a sunspot number. This was done just by calculating the number of spots present at any day where observations were undertaken. In order to be able to make a fair comparison to other records, we have added days, where it is stated in the notebooks that observations were made, but no sunspots were seen. We have however, not included sunspot observations where we cannot obtain complete heliographic coordinates directly. 

The sunspot number as a function of time is provided in Figure 1. The observations clearly show the indication of a solar cycles with a maximum in 1770. 
\begin{figure} 
	\centerline{\includegraphics[width=1.0\textwidth,clip=]{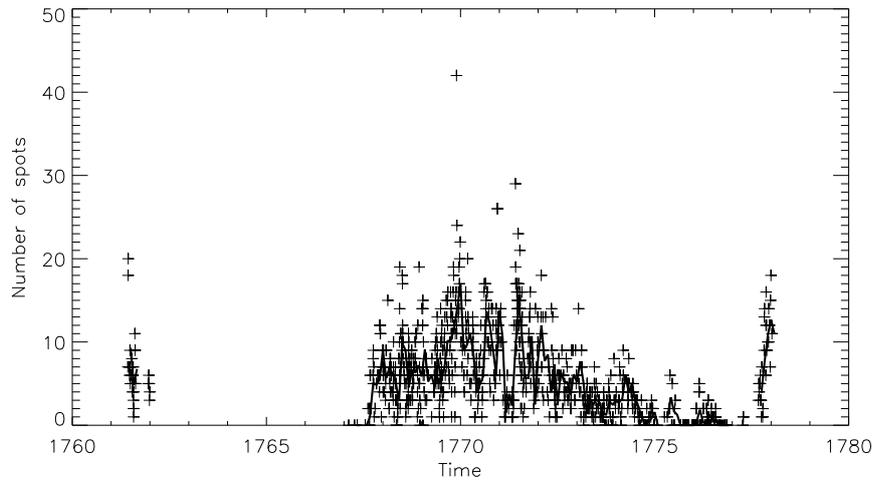}}
	\caption{Sunspot record reconstructed from Christian Horrebow's notebooks. The crosses are the daily measurements, whereas the solid line represents the monthly mean.}
\end{figure}
We have compared this sunspot record to three other records based on the observations by Christian Horrebow. The different records do not show perfect agreement, which is, at least partly, due to the fact that they do not measure the same thing. \citet{1995SoPh..160..387H} reconstructed only the number of sunspot groups, whereas d'Arrest and \citet{1859AN.....50..257T} reconstructed both sunspot groups and individual sunspots. Even though d'Arrest  and \citet{1859AN.....50..257T} measured the same quantity, groups, their reconstructions do also not show perfect agreement.

\begin{figure} 
	\centerline{\includegraphics[width=1.0\textwidth,clip=]{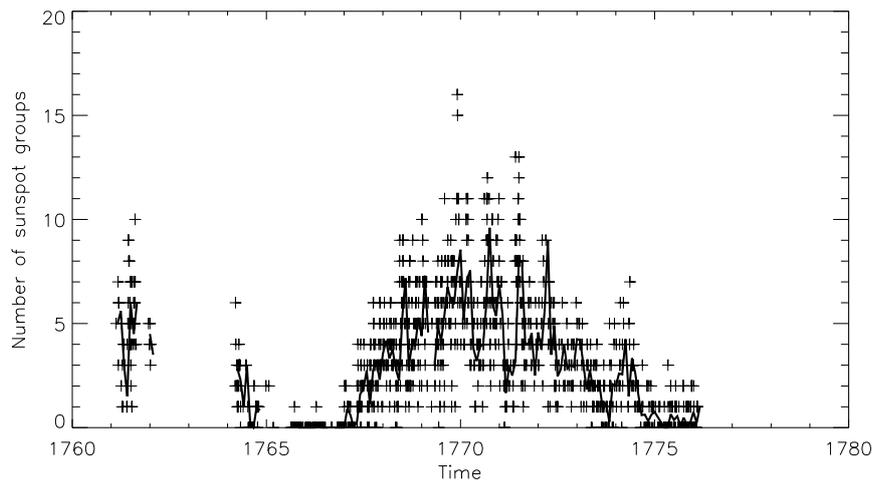}}
	\caption{Number of sunspot groups compiled by \citet{1995SoPh..160..387H}. The crosses are the daily measurements, whereas the solid line represents the monthly mean.}%\label{fig:?}
\end{figure}

The first comparison is made with the record compiled by \citet{1995SoPh..160..387H} based on the original notebooks, {\it i.e.} that same material used in this study. The sunspot number is calculated as the group sunspot number \citep{1998SoPh..181..491H}, however some differences can be expected between this and other records (see Figure~2). For this study we used the daily measurements that were available at the ftp server given in \citet{1998SoPh..181..491H}. Unfortunately, the link to the ftp server has not been active for some years. The measurements can now be found at the webpage of the National Geophysical Data Center\footnote{http://www.ngdc.noaa.gov/stp/space-weather/solar-data/solar-indices/sunspot-numbers/group/}.

\begin{figure} 
	\centerline{\includegraphics[width=1.0\textwidth,clip=]{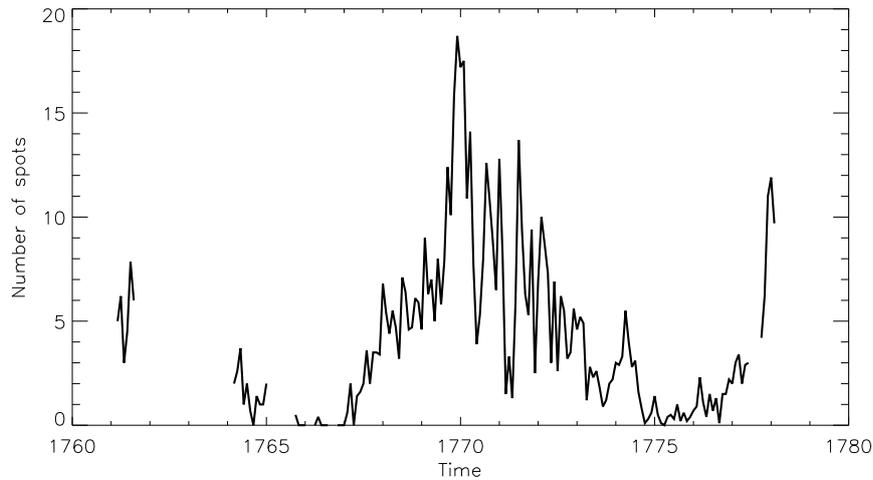}}
	\caption{Sunspot record compiled by \citet{1859AN.....50..257T}. The solid line represents the monthly mean.}
\end{figure}
\begin{figure} 
	\centerline{\includegraphics[width=1.0\textwidth,clip=]{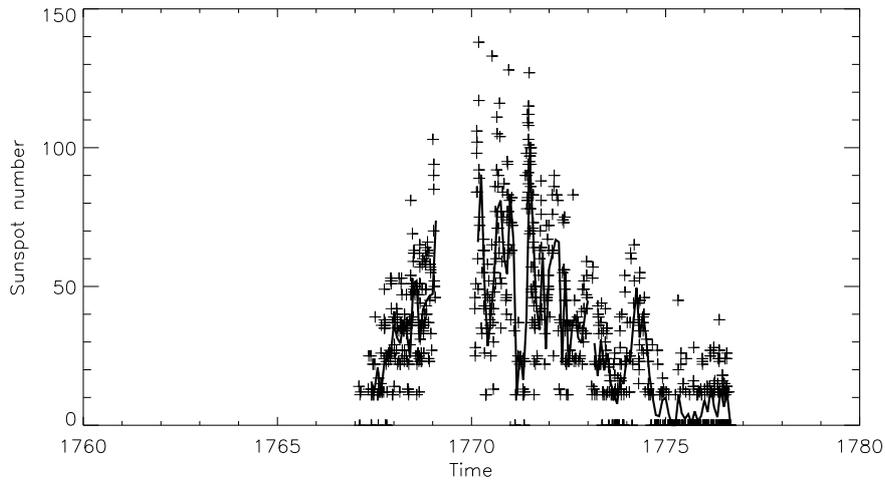}}
	\caption{Sunspot record compiled by d'Arrest. The solid line represents the monthly mean.}
\end{figure}
The second record containing sunspot monthly means was compiled by Thorvald Nicolai Thiele \citep{1859AN.....50..257T} (see Figure~3) and the third record was compiled by Heinrich Louis d'Arrest and published by \citet{1873MiZur...4...77W} (see Figure~4).

It appears that d'Arrest was not only the doctoral thesis supervisor of Thiele. Thiele also succeed d'Arrest in 1875 as professor at Copenhagen University and as director of the observatory. A position d'Arrest had held since 1857 \citep{Pedersen18}. Therefore, when Thiele published his record in 1859 as an astronomy student, he must have been one of the first students of d'Arrest. When d'Arrest later sent his record to Wolf in 1873, only two years before he died and Thiele took over his position, he must surely have known about Thiele's record and considered his record as an improvement. It is not clear why d'Arrest's records only contain sunspot observations from 1767 to 1776, but we speculate that this was done only to include the superior observations made with {\it Machina Æquatorea}.

The records by both \citet{1859AN.....50..257T} and d'Arrest gives both the number of sunspot groups and the number of individual sunspots. We have calculated the sunspot number using the following formulation:
\begin{equation}
R=\left( 10g + s \right),
\end{equation} 
where $R$ is the sunspot number, $g$ is the number of sunspot groups, and $s$ is the number of individual sunspots.

Caution should be taken when comparing the different sunspot reconstructions, as they were clearly not made using the same assumptions. Especially, definitions of what is an individual sunspot and what is a sunspot group changes between observers. Here it appears that \citet{1859AN.....50..257T} has a larger likelihood of calling something a sunspot group than d'Arrest had \citep{1995SoPh..160..387H}. Also, \citet{1995SoPh..160..387H} have only provided records of sunspot groups, whereas we have only provided records of individual sunspots. We have however, tried to compare the four different reconstructions. This has been done by binning the four reconstructions to the same monthly grid. This allowed us to both calculate the correlation between the different reconstructions and to plot the different reconstructions as a function of each other (Figures 5--10). In general, it is seen that the correlations are close to or above 0.9 between all reconstructions. The correlation between d'Arrest and \citet{1859AN.....50..257T} (Figure 10) is even above 0.95, which is not surprising given the close connection between the two. 

As expected the correlation analysis shows agreement, though not perfect, between the different reconstructions. We are not able to identify, based on the correlation analysis, if one of the reconstructions is better than the other. Our reconstruction will however, be the first one that will also provide sunspot positions.

\begin{figure} 
	\centerline{\includegraphics[width=1.0\textwidth,clip=]{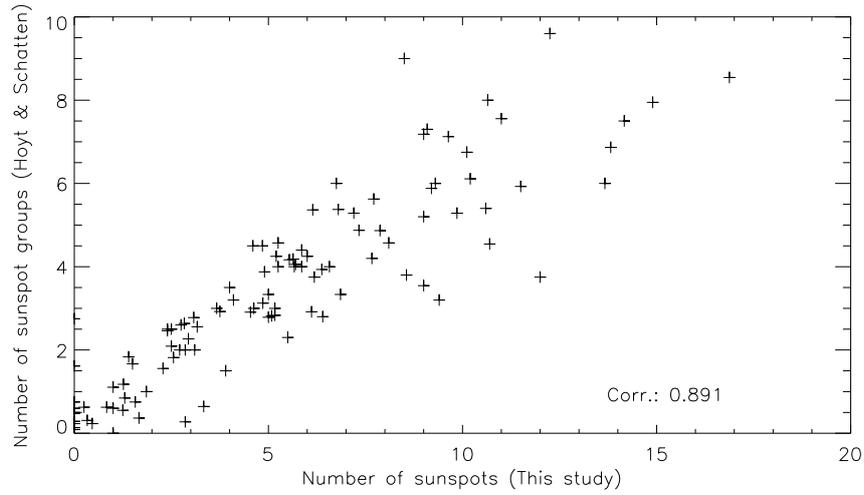}}
	\caption{Correlation between the sunspot reconstruction in this work and the reconstruction by \citet{1995SoPh..160..387H}.}%\label{fig:?}
\end{figure}
\begin{figure} 
	\centerline{\includegraphics[width=1.0\textwidth,clip=]{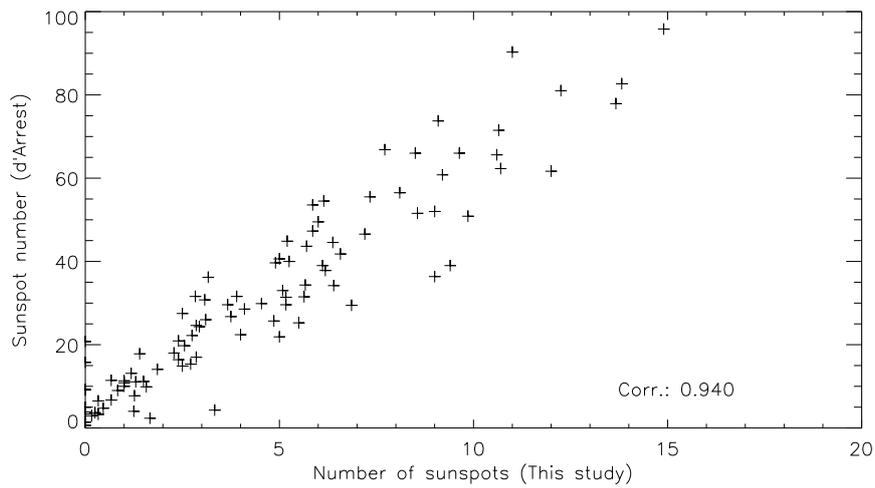}}
	\caption{Correlation between the sunspot reconstruction in this work and the reconstruction by d'Arrest.}%\label{fig:?}
\end{figure}
\begin{figure} 
	\centerline{\includegraphics[width=1.0\textwidth,clip=]{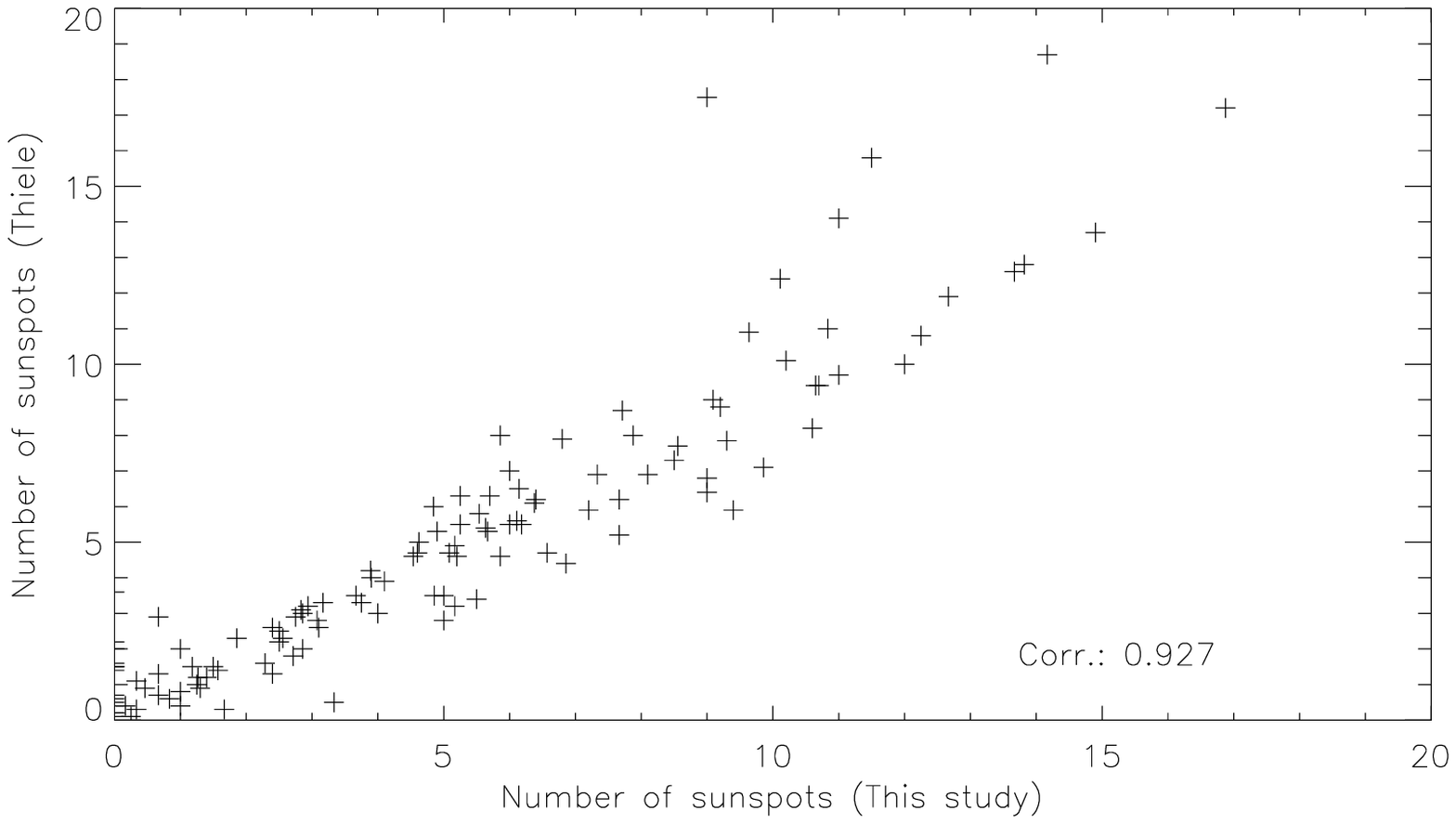}}
	\caption{Correlation between the sunspot reconstruction in this work and the reconstruction by \citet{1859AN.....50..257T}}%\label{fig:?}
\end{figure}
\begin{figure} 
	\centerline{\includegraphics[width=1.0\textwidth,clip=]{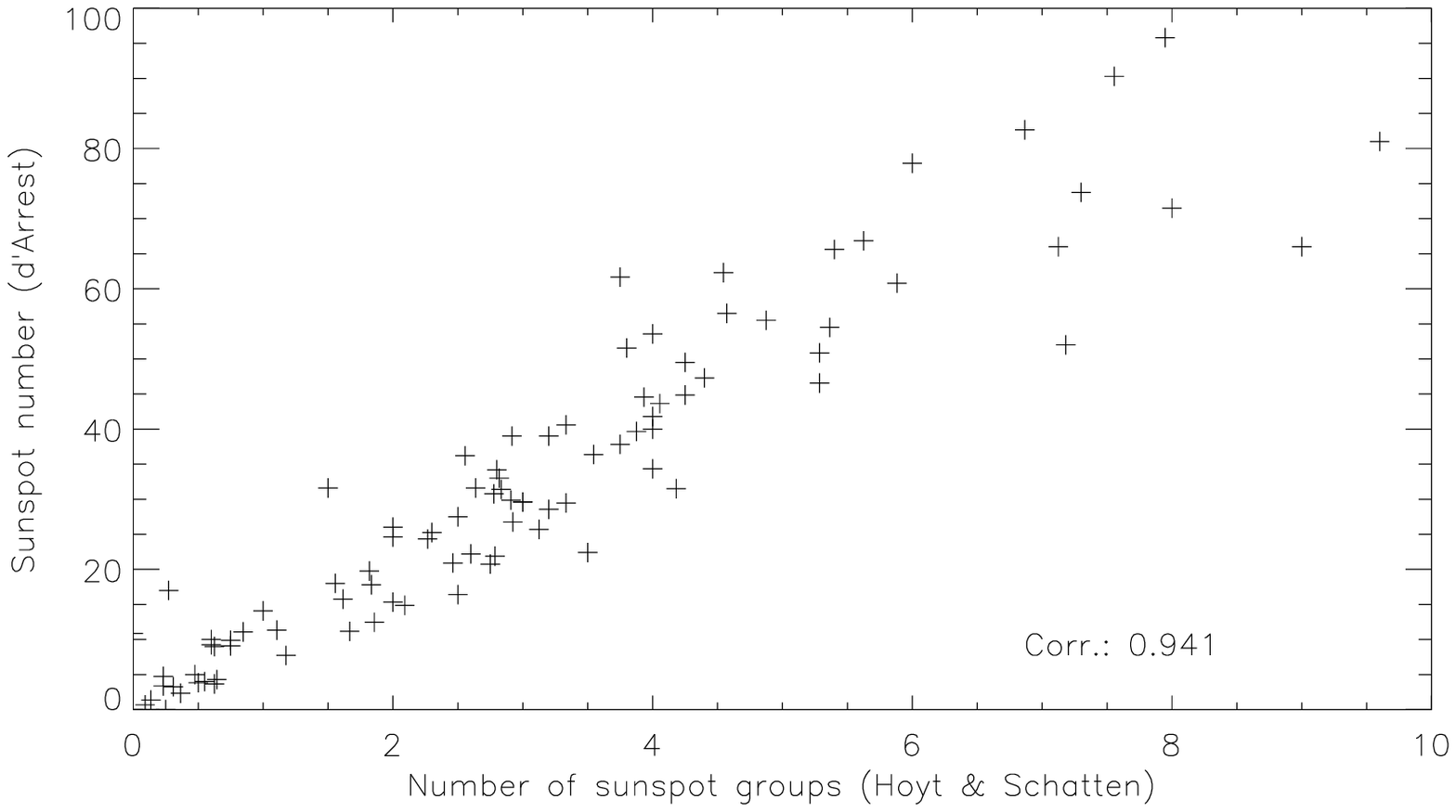}}
	\caption{Correlation between the sunspot reconstruction by \citep{1995SoPh..160..387H} and the reconstruction by d'Arrest.}%\label{fig:?}
\end{figure}
\begin{figure} 
	\centerline{\includegraphics[width=1.0\textwidth,clip=]{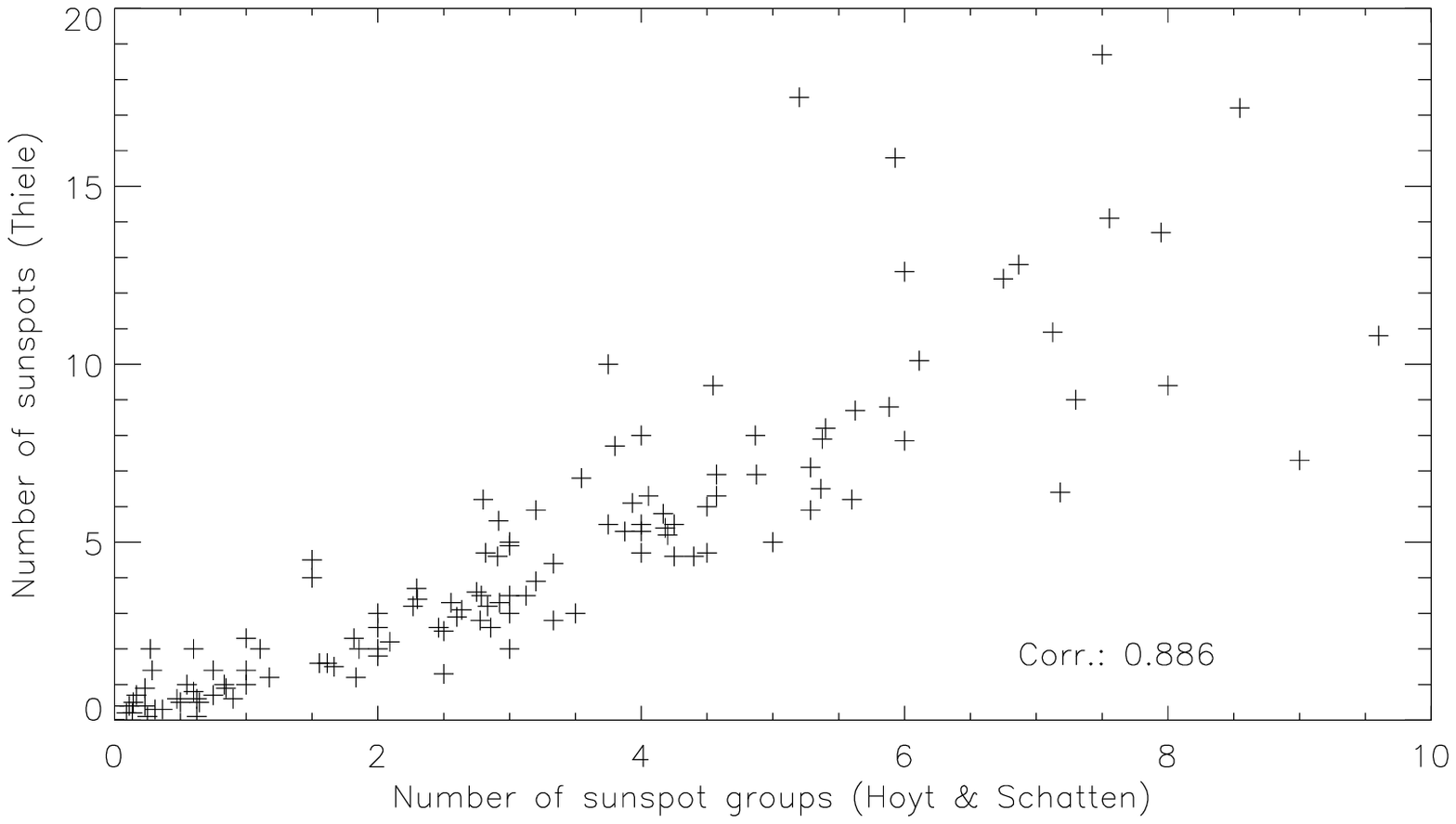}}
	\caption{Correlation between the sunspot reconstruction by \citep{1995SoPh..160..387H} and the reconstruction by \citet{1859AN.....50..257T}.}%\label{fig:?}
\end{figure}
\begin{figure} 
	\centerline{\includegraphics[width=1.0\textwidth,clip=]{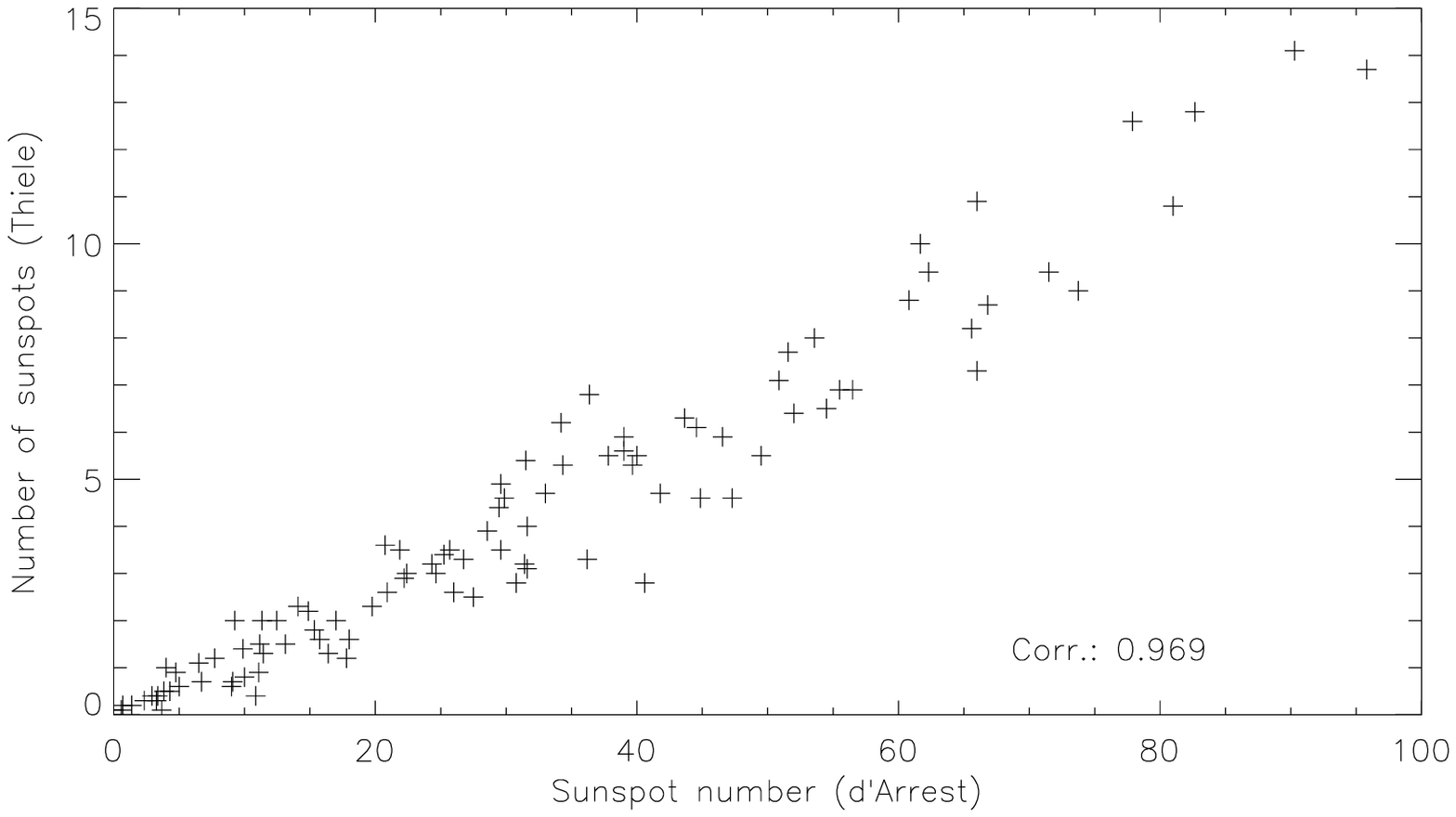}}
	\caption{Correlation between the sunspot reconstruction by d'Arrest and the reconstruction by \citet{1859AN.....50..257T}.}%\label{fig:?}
\end{figure}

\section{Butterfly Diagram}
In order to construct a butterfly diagram we need to transform the measured sunspot positions into heliographic coordinates. From the drawings, it is seen that the spots move from right to left over time, which suggests that the sunspot drawings are rotated 180 degrees with respect to the positions in the tables. There is however, no mention of any rotation of the images in the notebooks. By comparing the drawings with the positions in the tables, it is seen that the time of the leftmost spot is recorded first so the western limb is situated on the left. Before each vertical measurement, it is written from which limb, top or bottom, the vertical distance of spots is measured. The positions in the tables agree well with the alignment of limbs in the drawings. Hence, the western limb of the Sun is situated on the left whereas the eastern limb is situated on the right and the bottom limb is the northern limb. All sunspot drawings seem to have this orientation, except for a few drawings from July and November 1761 that are rotated by $180^\circ$ compared to the positions provided in the tables.

The sunspot observing times are given in local sidereal time. We checked the observing times by calculating the solar elevation angles for the corresponding times and found that the times of 35~observations were incorrect as they showed solar elevation angles of less than $0^\circ$. Most of these incorrect times referred to the times around midnight, that is why it is likely that these errors are due to confusion between a 12~h and 24~h time system. We thus added 12~h to these 35 observations and recalculated the solar elevation angle. Now, all observations of the Sun were above the horizon. Figure~11 shows the local hour angle, which is the right ascension subtracted from the local sidereal time, for the sidereal observing times of the eastern limb of the solar disc. The red boxes in the plot correspond to the observing times, which showed negative solar elevation angle before the addition of 12~h, and the green boxes correspond to the corrected sidereal times of observation with an addition of 12~h. The gray area in the plot corresponds to the time between sunset and sunrise. Hence, from Figure~11, it can be seen that the times of observation fall within the period of the Sun being above the horizon after the correction to the sidereal times. The corrected sidereal times are further used for calculation of heliographic coordinates of sunspots.

\begin{figure}[h]
\centering
\includegraphics[width=\textwidth]{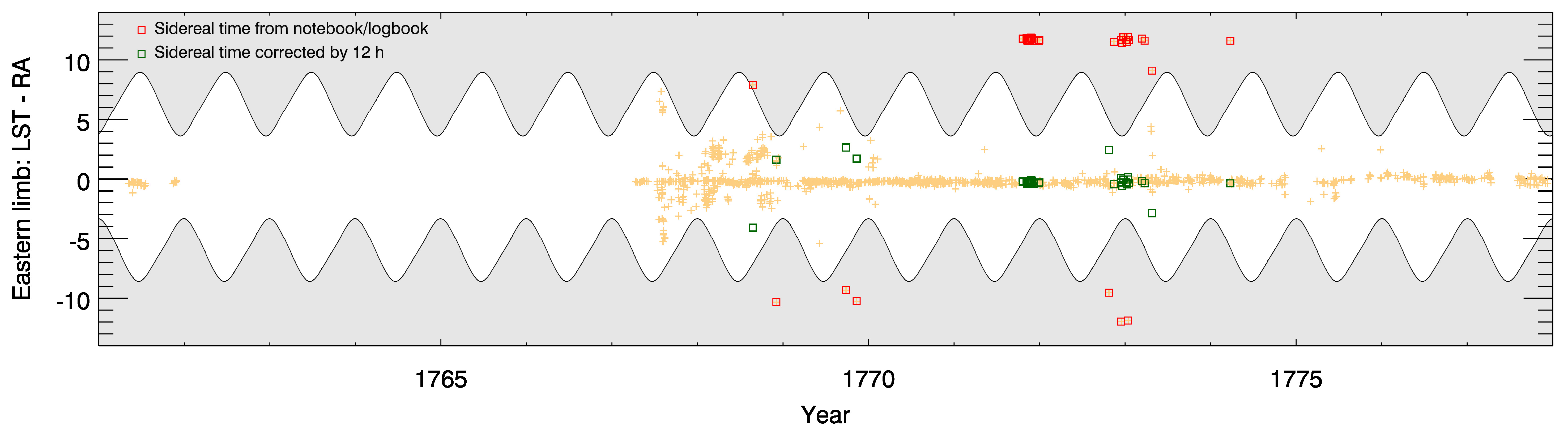}
\caption{The local hour angles for the sidereal times when the eastern limb of the Sun was observed. The gray region in the plot refers to the time between sunset and sunrise. The red boxes and green boxes refer to the sidereal times before and after the addition of 12~h to the sidereal observing times of the eastern limb, respectively, while the yellow crosses refer to the unaltered readings from the observation.
\label{hrbw_solelv}}
\end{figure}

The horizontal positions of sunspots are given in units of local sidereal time {\it i.e.}, the time between the observations of the western limb and the spot in minutes and seconds and the vertical distance of sunspots, either from the lower or the upper limb, are given in screw turns of the instrument used. For the vertical measurements made with {\it Machina Æquatorea} and {\it Machina Parallactican}, the distance of the spots is always measured from the bottom limb. For {\it Qvadrante} and {\it Rota Meridiana}, the distance of spots can be measured from either the top or the bottom limb. The orientation is, however, always given at the top of the vertical position readings. The horizontal and vertical diameters of the Sun are also given in the notebooks, either as transit times between passage of western and eastern limbs of the Sun or as screw turns. These diameters can be used to convert the horizontal and vertical sunspot positions from local sidereal time and screw turns into x and y coordinates of the spots. From the JPL HORIZONS ephemeris generator\footnote{https://ssd.jpl.nasa.gov/horizons.cgi}, we obtain ephemeris providing, the tip of the celestial equator and the tilt of the line of sight of the Sun's center in 6~h intervals at the geographical coordinates of Copenhagen observatory. From these ephemeris, we calculate the tip and the tilt for the times corresponding to the centre of the solar disc. 
The tilt of the solar disc is an important parameter in the calculation of latitudes of the sunspots, which are obtained from the vertical coordinates of the sunspots. The times at which vertical readings were measured are not given in the notebooks. However, in the observation description (Section~5.1 in Paper~I), it is mentioned that the horizontal coordinates of the sunspots were measured first and then the vertical coordinates, and so the nearest known time of eastern limb passage is considered for the calculation of the tilt. The heliographic grids with the obtained tilts and tips of the Sun are imposed on the x- and y-coordinates of sunspots, and then the heliographic latitudes and the distances from the central meridian of the solar disc are obtained. The longitudes of the solar disc centre, corresponding to the sidereal times of disc centre, are obtained from the ephemeris and then the heliographic longitudes of sunspots are calculated using the longitudes of the solar disc centre and the central meridian distances of sunspots.

\begin{figure}[h]
\centering
\includegraphics[width=\textwidth]{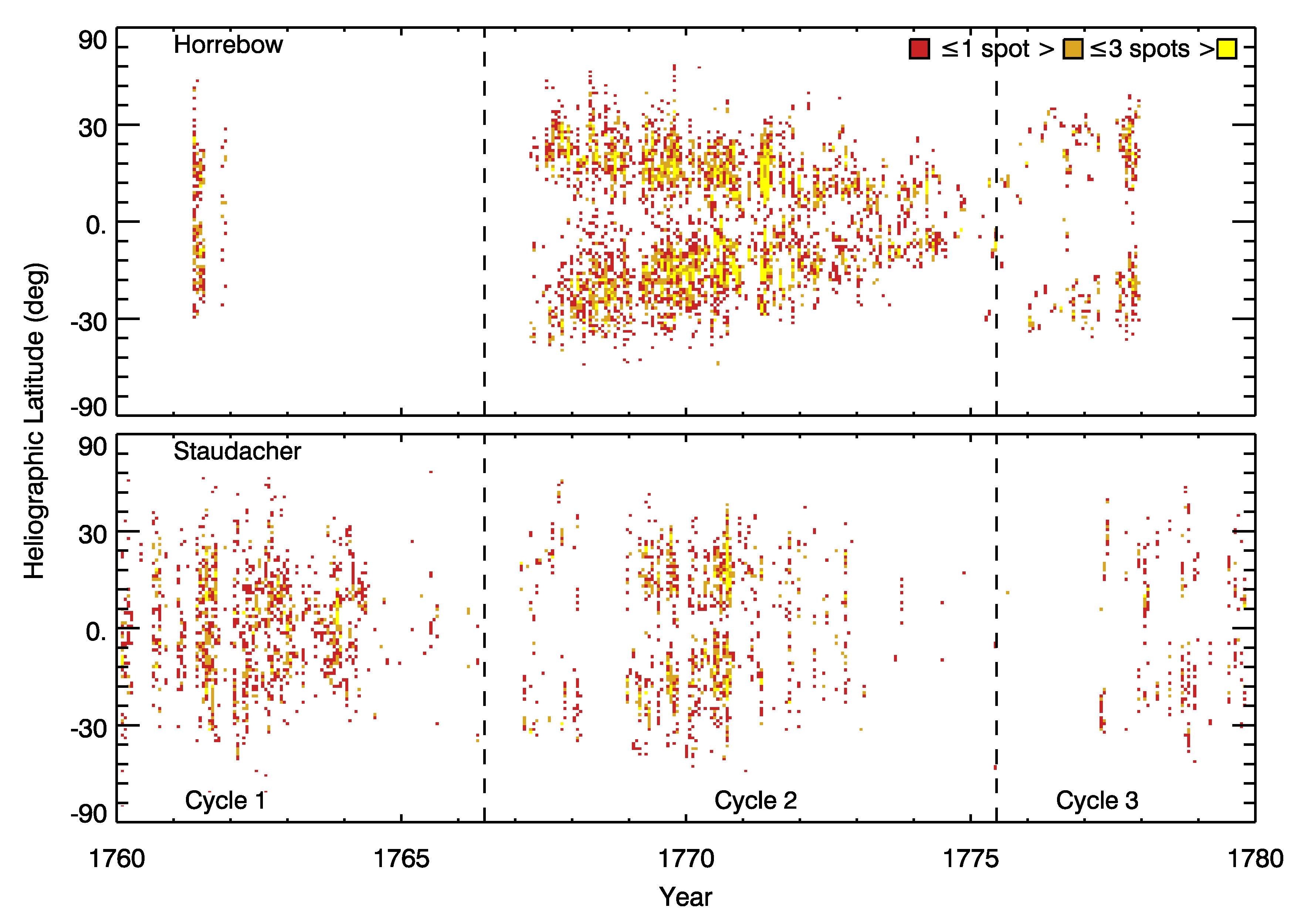}
\caption{Top panel: Butterfly diagram obtained from Christian Horrebow's observations for the solar cycles~1--3. Bottom panel: comparison to the butterfly diagram obtained from Staudacher's observations. The dashed lines denote the times of minima for solar cycles~2 and 3, which are taken from the ``Average'' column of the cycle timings by \cite{2015LRSP...12....4H}.
\label{hrbw_butterfly}}
\end{figure}

Some of the numbers in the tables are follow by a comment: "$.$", "$-$", "$v$" and "$d$". This is also the case for the tables in Horrebow's printet articles. In {\it Translation of Christian Horrebow's writings in Dansk Historisk Almanak 1770} it is explained that "$.$" means that one-fourth should be added to the number, "$-$" means add one-half and "$v$" means that one-fourth should be subtracted. "$d$" on the other hand means that the number is uncertain. We have not applied these corrections to the numbers. The main reason for this is that the correction can be hard to read, especially the difference between "$.$" and "$-$". Also, a correction of half a second corresponds to less than half a degree in heliographic coordinates at solar disc center, which is far below the precision for these observations.

We obtained unique positions of 6648~spots from the notebooks for 1020~days during the years 1761, 1767--1777. Some days include more than one observation and in total we therefore have 1171 observations. The heliographic coordinates of sunspots are obtained for 6622~spots, whereas the remaining 26~spots lay outside the solar disc and so the heliographic coordinates could not be measured. Figure~12 shows the butterfly diagram calculated based on Christian Horrebow's sunspot observations. This is also compared to the butterfly diagram calculated based on Staudacher's sunspot observations. If there are more than one observation of a sunspot in a day, then the average of heliographic latitudes of a sunspot is used for the butterfly diagram which reduces the total number of sunspots from 6622 to 6170. The calculation of the average latitude of a sunspot is possible as the data contains label for each spot. Figure~12 shows the butterfly diagram obtained from 6170~spots of Horrebow data which covers a year during the maximum phase of solar cycle~1, full solar Cycle~2 except around a year in the beginning, and the beginning of Solar Cycle~3. 

\section{Discussion}
The reanalysis of Horrebow's notebooks have shown us that sunspot positions and not just sunspot numbers can be extracted from the observations. The fact that we see very few sunspots on or near the solar equator during Cycle 2 indicates that sunspot positions have a reasonable precision. 

In general our result supports the conclusion by \citet{2009SoPh..255..143A}, based on observations by Staudacher, that Cycle 0 and 1 contained a significant number of sunspots on or near the solar equator, while Cycle 2, 3, and 4 did not, though this is mainly based on the observations in 1761. Our measurements do therefore not contradict the suggestion by \citet{2009SoPh..255..143A} of the presence of a dynamo mode that is symmetric around the equator between the Maunder and Dalton minima. Future studies of i.e. tilt angles will allow us to test this hypothesis even further.

The aim of this study has not been to reconstruct the most reliable sunspot number record and in comparison with  other sunspot number reconstructions, ours does not also stand out. In fact one can argue that our reconstruction is less reliable as we do not reconstruct the group sunspot number. We do, however, speculate that the sunspot positions can be used to calculate a group sunspot number in an objective and transparent way. Such a procedure would have the advantage that it could provide not only the group sunspot number, but also the tilt angles of the sunspot groups. The tilt angles of the sunspot groups have lately received renewed attention with the growing interest in flux transport dynamo models \citep[see {\it i.e.}][]{1999ApJ...518..508D}. We have earlier used similar historic observations to reconstruct tilt angles of sunspot groups \citep{2015A&A...584A..73S, 2016AdSpR..58.1468S, 2016A&A...595A.104A} and intend to do the same for the Horrebow observations in a forthcoming paper (Senthamizh Pavai in preparation).

\section{Conclusion}
In this article we present a reanalysis of the sunspot observations by Christian Horrebow covering the period 1761 and 1764--1777. Contrary to an earlier analysis of Christian Horrebow's observations we also reconstruct the sunspot positions, which allows us to calculate a butterfly diagram covering the maximum phase of Cycle~1, full Cycle~2 except around a year in the beginning, and the beginning of Cycle~3. The butterfly diagram is in agreement with similar observations by Staudacher  \citep{2009SoPh..255..143A} that Cycle 0 and 1 did contain sunspot on or near the solar equator, while Cycles 2, 3 and 4 did not.

This article is the second one in our series on Christian Horrebow's observations of the Sun. Paper I contains a biography and a complete bibliography of Christian Horrebow and we expect a forthcoming article that will present tilt angles of the sunspot groups observed by Christian Horrebow (Senthamizh Pavai in preparation).

In connection with this paper, a version of Horrebow's notebooks is made available as electronic supplementary material.

%%%%%%%%%%%%%%%%%%%%%%%%%%%%%%%%%%%%%%%%%%%%%%%%%%%
%% Sections
%
% \section{}%\label{s:?} 

%% Figure 
%
% \begin{figure} 
% \centerline{\includegraphics[width=0.5\textwidth,clip=]{<fig.eps>}}
% \caption{}%\label{fig:?}
% \end{figure}

%% Table
%
% \begin{table}
% \caption{}%\label{tbl:?}
% \begin{tabular}{}     
% \hline
% \multicolumn{2}{c}{<>}
% <data>
% \hline
% \end{tabular}
% \end{table}

%%%%%%%%%%%%%%%%%%%%%%%%%%%%%%%%%%%%%%%%%%%%%%%%%%%%%%%%%%%%%%%%%%%%%%%%%%%
%% Appendix
%
% \appendix   

%%%%%%%%%%%%%%%%%%%%%%%%%%%%%%%%%%%%%%%%%%%%%%%%%%%%%%%%%%%%%%%%%%%%%%%%%%%
%% Acknowledgements
%
\begin{acks}
We would like to thank the referee for thoughtful comments, which significantly improved the paper. The project has been supported by the Villum Foundation. Funding for the Stellar Astrophysics Centre is provided by the Danish National Research Foundation (grant agreement no. DNRF106). We are very thankful to librarian Susanne Elisabeth Nørskov, who have been very helpful throughout the projects with digging out the notebooks and other historical texts. We are also thankful to Kristian Hvidtfelt Nielsen for help with interpreting the historical texts.
\end{acks}

%%% %%%%%%%%%%%%%%%%%%%%%%%%%%%%%%%%%%%%%%%%%%%%%%%%%%%%%%%%%%%
%% Bibliography
%
% Using BibTeX
%
% \bibliographystyle{spr-mp-sola}
% \bibliography{<bib file>}  
%
% Without BibTeX 

\end{article} 
\end{document}